\documentclass[prl,aps,twocolumn]{revtex4}

\usepackage{graphicx}
\usepackage{amsmath,empheq}
\usepackage{amssymb}
\usepackage{ifthen}
\usepackage{booktabs}
\usepackage{color}

\usepackage{graphicx}% Include figure files
\usepackage{dcolumn}% Align table columns on decimal point
\usepackage{bm}% bold math
\usepackage{longtable}

\newcommand{\bb}{\begin{equation}}
\newcommand{\ee}{\end{equation}}
\newcommand{\ba}{\begin{eqnarray*}}
\newcommand{\ea}{\end{eqnarray*}}
\newcommand{\rhor}{\rho({\bf r})}

\newcommand{\rr}{{\mathbf r}}
\newcommand{\dr}{{\rm d}{\bf r}}

\begin{document}

\title{Kelvin equation for bridging transitions}

\author{Alexandr \surname{Malijevsk\'y}}
\affiliation{{Department of Physical Chemistry, University of Chemical Technology Prague, Praha 6, 166 28, Czech Republic;}
 {The Czech Academy of Sciences, Institute of Chemical Process Fundamentals,  Department of Molecular Modelling, 165 02 Prague, Czech Republic}}
 \author{Martin \surname{Posp\'\i\v sil}}
 \affiliation{{Department of Physical Chemistry, University of Chemical Technology Prague, Praha 6, 166 28, Czech Republic}}

\begin{abstract}

\noindent We study bridging transitions between a pair of non-planar surfaces. We show that the transition can be described using a generalized Kelvin equation by mapping the system to a slit of
finite length. The proposed equation is applied to analyze the asymptotic behaviour of the growth of the bridging film, which occurs when the confining walls are gradually flattened. This phenomenon
is characterized by a power-law divergence with geometry-dependent critical exponents that we determine for a wide class of walls' geometries. In particular, for a linear-wedge model,  a covariance
law revealing a relation between a geometric and Young's contact angle is presented. These predictions are shown to be fully in line with the numerical results obtained from a microscopic (classical)
density functional theory.
\end{abstract}

\maketitle

It is widely recognized that confining a  fluid may dramatically change its phase behaviour \cite{RW,henderson92,binder2008}. Perhaps the most familiar example of this is the phenomenon of capillary
condensation which refers to a shift of the liquid-gas phase boundary when a fluid is confined between two parallel walls a distance $L$ apart \cite{NF81, NF83, parry90, gelb}. According to the
Kelvin equation, the condensation of the confined fluid occurs at a chemical potential $\mu$, which is shifted from the saturation line $\mu_{\rm sat}(T)$ by an amount \cite{gregg}
 \bb
 \delta\mu_{\rm cc}^{\rm slit}(L)=\frac{2\gamma\cos\theta}{L\Delta\rho}\,, \label{kelvin}
 \ee
where $\gamma$ is the liquid-gas surface tension, $\theta$ is Young's contact angle characterizing the wetting properties of the walls and $\Delta\rho=\rho_l-\rho_g$ is a difference between the bulk
liquid and gas number densities. Despite its simple form and macroscopic origin, the Kelvin equation is remarkably accurate even at the nanoscale where we can expect that finite-size effects are
particularly significant \cite{evans84, evans85, evans90}.

It is also well known that further and \emph{qualitatively} new phenomena may occur, if the planar symmetry of the confined system is broken \cite{rocken96, rejmer99, parry99, quere02, bruschi02,
gang, tas07, quere08, bruschi09, hofmann, mistura13, pos22}. In this case, different wall shapes may induce new types of phase transition whose nature is steered by both thermodynamic and geometric
parameters. For instance, the simple modification of making just one of the confining walls of finite extent $H$, results in remarkably complex behaviour from the interplay between two types of
capillary condensation, meniscus depinning transitions and corner filling phase transitions, which are controlled by the aspect ratio $a=L/H$ \cite{mal21}. Central to understanding this phenomena is
a concept of an \emph{edge contact angle} $\theta_e$, proposed originally for the description of capillary condensation inside a finite slit \cite{our_slit}. Here, $\theta_e$ is the angle at which
the menisci, formed in the condensed state and pinned at the walls' edges, meet the walls. Simple geometric arguments then dictate that condensation in the finite slit occurs at the chemical
potential $\mu_{\rm cc}^H=\mu_{\rm sat}-\delta\mu_{\rm cc}^H$, with \cite{our_slit}
 \bb
 \delta\mu_{\rm cc}^H=\frac{2\gamma\cos\theta_e}{D\Delta\rho} \label{kelvin_fs}
 \ee
where $H$ is the length of the walls and $D$ is the width of the slit (intentionally distinguished from $L$ for the further purposes). Right at $\mu_{\rm cc}^H$, the edge contact angle is given
implicitly by the equation
 \bb
 \cos\theta_e=\cos\theta-\frac{D}{2H}\left[\sin\theta_e+\left(\frac{\pi}{2}-\theta_e\right)\sec\theta_e\right]\,, \label{thetae}
 \ee
from which it follows that $\theta_e\to\theta^+$, as $H\to\infty$.

In this paper we show that the concept of an edge contact angle can be advantageously used for the description of another type of condensation this time induced by walls of arbitrary geometry and in
the absence of pinning. To this end, consider a pair of symmetric walls, such that the local height of the ``top'' (``bottom'') wall, relative to the horizontal plane $z=0$, is $z_w(x)>0$
($-z_w(x)$), which we express as:
 \bb
  z_w(x)=\frac{L}{2}+ \psi(x)\,. \label{zw}
 \ee
Here, $\psi(x)$ is assumed to be a differentiable function (except for $x=0$ where we allow for a possible kink) describing the shape of the walls that are of macroscopic extent along the remaining
$y$-axis.  We will further assume that $\psi(x)$ is even and, without any loss of generality, has its global minimum at the origin. We note that these assumptions are not crucial and can be easily
generalized, as will be discussed at the end.

\begin{figure*}
 \includegraphics[width=16cm]{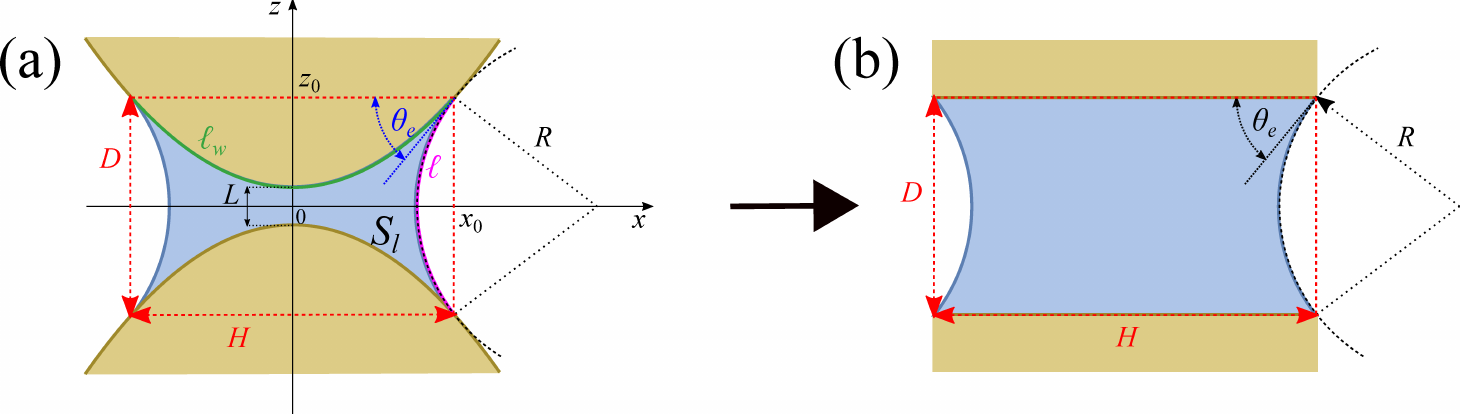}
\caption{Schematic illustration of a locally condensed state induced by a pair of non-planar walls (with location described by the function $z_w(x)$)  that are completely wet (panel a).
Macroscopically, the configuration is characterized by a presence of two symmetric menisci of Laplace radius $R=\gamma/(\delta\mu\Delta\rho)$ that connect the walls tangentially at the points $[\pm
x_0,\pm z_0]$. The length of contact between the liquid (occupying the area $S_l$) and each surface is $\ell_w$. To determine the location of the local condensation (bridging), the system can be
mapped onto a model system corresponding to a finite planar slit (panel b) with the parameters $H=2x_0$, $D=2z_0$ and the edge contact angle  satisfying  $\tan\theta_e=z_w'(x_0)$}.\label{fig_sketch}
\end{figure*}

If the contact angle of the walls is $\theta<\pi/2$ ($\theta>\pi/2$), the system exhibits local condensation (evaporation), i.e. a \emph{bridging transition}, near the origin when the chemical
potential $\mu_B=\mu_{\rm sat}-\delta\mu_B$, which can be determined by mapping the system to that of a finite slit, as illustrated in Fig.~\ref{fig_sketch}  (for completely wet walls). In this way
it follows that the condition for the bridging transition can be formally expressed in the same way as for condensation in a finite slit
 \bb
 \delta\mu_B=\frac{2\gamma\cos\theta_e}{D\Delta\rho}\,, \label{bridging}
 \ee
in the following sense. Let $[x_0, z_0\equiv z_w(x_0)]$ denote the point of contact between the meniscus and the wall (in the first quadrant). We now associate the bridged part of our system with a
finite slit of length $H=2x_0$ and width $D=2z_0$, while the corresponding edge contact angle is determined by the slope of the wall tangent at $x_0$, $\theta_e=\tan^{-1}(z_w'(x_0))+\theta$. At the
bridging transition, the Laplace radius of the meniscus is $R=\gamma/(\delta\mu_B\Delta\rho)$, giving the geometric condition Eq.~(\ref{bridging}), valid for any bridged (locally condensed) state.
The free energy balance between the unbridged and bridged states dictates that the value of $\theta_e$ corresponding to the bridging transition, is given by
 \bb
 \bar{r}\cos\theta_e=r\cos\theta-\frac{D}{2H}\left[\sin\theta_e+\left(\frac{\pi}{2}-\theta_e\right)\sec\theta_e\right]\,, \label{thetae2}
 \ee
which extends Eq.~(\ref{thetae}) by introducing two dimensionless parameters: the ``roughness''  $r=\ell_w/H$ and its two-dimensional analogue $\bar{r}=S/(HD)$. Here,
$\ell_w=2\int_0^{x_0}\sqrt{1+(\psi'(x))^2}dx$ is the liquid-wall contact length and $S=4\int_0^{x_0}z_w(x)dx$ corresponds to the available volume between the confining walls, which consists of the
portion occupied by liquid, $S_l=S-S_g$ and the portion filled by gas $S_g=(\pi-2\theta_e)R^2-\sin\theta_e RD$ (see Fig.~1a). We note that the bridging transition corresponds to local condensation
provided the aspect ratio $D/H<\cos\theta^*$ where $\theta^*$ can be interpreted as the apparent contact angle given by Wenzel's law, $\cos\theta^*=r\cos\theta$ \cite{wenzel}. Similarly for
$D/H>\cos\theta^*$, $\theta_e>\pi/2$ the bridging transition corresponds to a local evaporation (with bulk phase being a liquid).

At this point, we make several pertinent remarks regarding the applicability and limitations of Eq.~(\ref{bridging}).
 %Just as its original version for condensation in infinite slits, the modified
%Kelvin equation is macroscopic in nature and cannot thus account for interfacial phenomena driven by microscopic forces, which may somewhat shift the bridging phase boundary.
Specifically, in the derivation of the modified Kelvin equation we did not consider related interfacial phenomena, which under certain conditions may also occur.
 For instance, for walls possessing pockets or grooves, the individual walls may experience filling or unbending transitions below $T_w$ accompanied by a jump in adsorption at the
troughs  \cite{unbending, manuel}. Since this will effectively reduce both $\bar{r}$ and $r$, one may anticipate that the opposing effects in the change of both parameters may to some extent
compensate; however, further numerical tests are needed. Above $T_w$, wetting layers will adsorb even at weakly corrugated walls which will primarily affect the parameter $\bar{r}$ and shift the
transition closer to saturation. This effect could be incorporated to Eq.~(\ref{bridging}) using the construction of Rasc\'on and Parry \cite{rascon2000}, as also demonstrated recently for sinusoidal
walls \cite{posp_22}. Of course, packing effects, that are particularly significant near the surface of highly curved walls, may also affect the bridging scenario, as well as bulk critical phenomena
which would not allow for bridging transition if the walls' separation becomes comparable with the bulk correlation length. Here, our main focus is on analyzing asymptotic behaviour of bridging
transitions for some simple, yet important model geometries utilizing the modified Kelvin equation and support the outcomes by comparing the analytic predictions with the numerical results using a
more microscopic approach.

In the remaining part of our paper we illustrate the utility of the generalized Kelvin equation (\ref{bridging}) emphasizing the connection between bridging and capillary condensation. We show that
for a wide range of wall geometries, the local condensation between walls generates a sort of critical phenomenon associated with the divergence of condensed film thickness, with geometrically
dependent critical exponents, when the confining walls are flattened. Finally we test the analytic predictions on a microscopic level using a classical non-local density functional theory (DFT).

 \begin{figure*}
 \includegraphics[width=\linewidth]{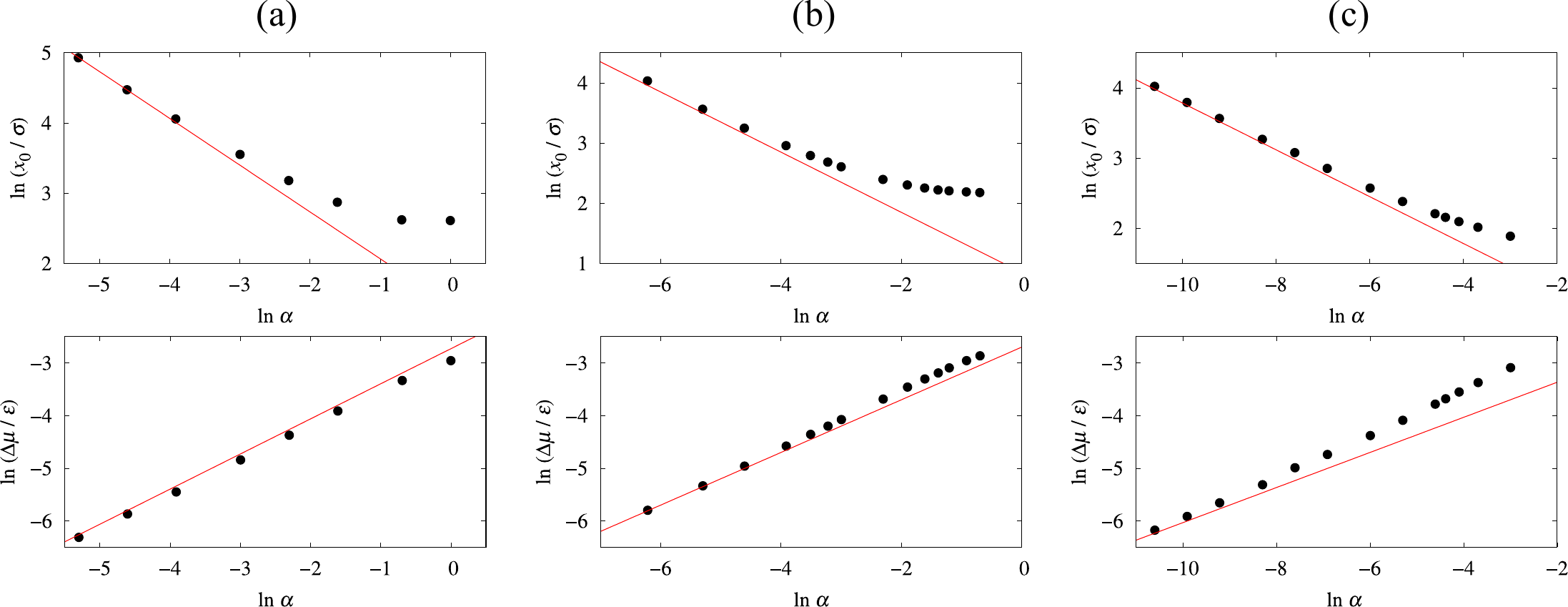}
\caption{Log-log plots showing the growth of the bridging films $x_0$ (upper panels) and the decrease of $\Delta\mu=\mu_B-\mu^{\rm slit}_{\rm cc}$ (lower panels) for the power-law shaped walls given
by Eq.~(\ref{gen_wall}) with a) $\nu=1/2$, b) $\nu=1$, and c) $\nu=2$ upon reducing the parameter $\alpha$. The symbols represent the DFT results, while the straight lines have the slopes
corresponding to the expected values of the critical exponent $\beta_\alpha=1/(1+\nu)$. In all the cases the minimum distance between the walls is $L=10\,\sigma$.} \label{fig2}
\end{figure*}

We illustrate the use of Kelvin's equation for bridging transitions between walls whose shape is described by the power law
 \bb
  \psi(x)=\alpha\frac{|x|^\nu}{L^{\nu-1}}\,,\label{gen_wall}
 \ee
with $\alpha,\nu>0$, as originally suggested by Rasc\'on and Parry \cite{rascon2000} in their study of adsorption at a single wall. Hereafter, we will focus on the case of completely wet walls
($\theta=0$), for which $\cos\theta_e=1/\sqrt{1+\psi'^2(x_0)}$, where $x_0$ is given by Eq.~(\ref{thetae2}). Clearly, it generally holds that $\delta\mu_B(L)<\delta\mu^{\rm slit}_{\rm cc}(L)$, but as
the steepness parameter $\alpha$ tends to zero, so does the difference $\Delta\mu=\mu_B-\mu^{\rm slit}_{\rm cc}$ according to the power-law
 \bb
 \Delta\mu\sim\delta\mu^{\rm slit}_{\rm cc}(L)\alpha^{\beta_\alpha}\,,\;\;\alpha\to0\,,\label{dmu_genwall}
 \ee
 where  $\beta_\alpha=1/(1+\nu)$.
The process of the walls' flattening is accompanied by a growth of the bridging film thickness, with the asymptotic behaviour
 \bb
 x_0\sim L\alpha^{-\beta_\alpha}\,,\;\;\alpha\to0\,,\label{x0_genwall}
 \ee
with  a positive subdominant contribution of the order of ${\cal{O}}(1)$. From the asymptotic behaviour of $\Delta\mu$ and $x_0$ it follows that for steeper walls (large $\nu$), the bridging
transition occurs further away from $\mu^{\rm slit}_{\rm cc}$ and the growth of the bridging film is slower. In contrast to $x_0$,  the sub-dominant correction to  $\Delta\mu$ strongly depends on the
walls geometry, and in particular on their curvature, as follows:

\noindent {\bf i) $\nu>1$}: For convex walls (positive curvature), the subdominant contribution to $\Delta\mu$ is of the order of ${\cal{O}}\left(\alpha^{\frac{2}{\nu+1}}\right)$
which is \emph{positive} meaning that the asymptotic result, Eq.~(\ref{dmu_genwall}), approaches the exact solution from below.\\
\noindent {\bf ii) $\nu<1$}: For concave walls (negative curvature), the subdominant contribution to $\Delta\mu$ is of the order of ${\cal{O}}\left(\alpha^{\frac{2}{\nu+1}}\right)$
which is \emph{negative} meaning that the asymptotic result, Eq.~(\ref{dmu_genwall}), approaches the exact solution from above.\\
\noindent {\bf iii) $\nu=1$}: In the marginal, linear case (zero curvature), the subdominant contribution to $\Delta\mu$ is   only ${\cal{O}}\left(\alpha^{\frac{4}{\nu+1}}\right)$ and is positive.

The linear, double-wedge model, $\nu=1$, turns out to be specific also for other reasons. Firstly, it allows for the exact explicit solution of the Kelvin equation for any value of $\alpha$, which
can be expressed as

% \bb
% \frac{x_0}{L}=\frac{\xi\phi+\sqrt{2\xi(2\alpha+\phi)}}{2\xi(2-\alpha\phi)}\,, \label{x0_wedge}
% \ee
%where $\phi=\pi-2\tan^{-1}\alpha$ and $\xi=\alpha+\alpha^3$. Secondly, there exists a ``covariance law''
% \bb
% \delta\mu_B^\alpha(L;\theta=0)=\delta\mu_{\rm cc}(\tilde{L};\theta=\tilde{\alpha})\,, \label{cov}
% \ee
% where $\tilde{\alpha}=\tan^{-1}(\alpha)$ is a tilt angle (relative to the horizontal) of the walls and
% \bb
%\tilde{L}(\alpha)=L+2x_0\alpha \,,\label{leff}
% \ee
% which for sufficiently small values of $\alpha$ becomes
% \bb
% \tilde{L}(\alpha)\approx L(1+\sqrt{\pi\alpha/2})\,, \label{leff_asm}
% \ee
%relating bridging transition induced by completely wet wedges of a tilt angle $\tilde{\alpha}$  and capillary condensation in infinite planar slit formed of partially wet walls with the contact angle
%$\theta=\tilde{\alpha}$.

 \bb
 \frac{x_0}{L}=\frac{\xi\phi+\sqrt{2\xi(2\alpha+\phi)}}{2\xi(2-\alpha\phi)}\,, \label{x0_wedge}
 \ee
where $\phi=\pi-2\tan^{-1}\alpha$ and $\xi=\alpha+\alpha^3$. Secondly, there exists a \emph{covariance law}
 \bb
 \delta\mu_B^\alpha(L;\theta=0)=\delta\mu^{\rm slit}_{\rm cc}(\tilde{L};\theta=\tan^{-1}(\alpha))\,, \label{cov}
 \ee
 where
 \bb
\tilde{L}(\alpha)=L+2x_0\alpha \,.\label{leff}
 \ee
 For sufficiently small values of $\alpha$, the parameter plays a role of a tilt angle (relative to horizontal) and the covariance law simplifies, such that
 \bb
 \tilde{L}(\alpha)\approx L(1+\sqrt{\pi\alpha/2})\,, \label{leff_asm}
 \ee
relating bridging transition induced by completely wet wedges of a tilt angle $\alpha$  and capillary condensation in infinite planar slit formed of partially wet walls with the contact angle
$\theta=\alpha$.

%and in particular for $\nu=1$ it holds for sufficiently small values of $\alpha$ that

% \bb

%\delta\mu_B(L;0)\approx\delta\mu{\rm cc}(L;\alpha)\left(1-\sqrt{\frac{\pi}{2}\alpha}\right)\,.

% \ee

%This is an example of a covariance law, which tights bridging transition between completely wet walls with capillary condensation inside a slit

%formed of partially wet walls, by associating thermodynamic contact angle with the geometric parameter of the model.

The results (\ref{dmu_genwall}) and (\ref{x0_genwall}) are rather general and applicable for a wide range of confinement models. This can be illustrated by considering a pair of walls of circular
intersections of radius $r$ with $\psi(x)=r-\sqrt{r^2-x^2}$, corresponding to a pair of discs in 2D and to a pair of parallel cylinders in 3D. Here, the large $r$ analysis of the bridging transition
leads to the results

% In this case, the Kelvin equation (\ref{bridging}) can be recast into the form of

 %\begin{eqnarray}

%&&(L+2r)x_0+(R^2-r^2)a\sqrt{1-a^2}+R^2\cos^{-1}a\nonumber\\

%&&-(2R+r)r\sin^{-1}a=0\,, \label{cyl_kelvin}

% \end{eqnarray}

% where $a=x_0/r$ and $R$ is the Laplace radius of the bridging menisci. Focusing on the limit of large values of $r$, it follows that [SM]

 \bb
 \delta\mu_B=\delta\mu^{\rm slit}_{\rm cc}(L)\left[1-c^2\left(\frac{L}{r}\right)^{\frac{1}{3}}\right]\,,\;\;\;r\gg L\label{dmu_cyl}
 \ee
 and
 \bb
 x_0=c(L^2r)^{\frac{1}{3}}\,,\;\;\;r\gg L\,,\label{x0_cyl}
 \ee
with $c=(3\pi/16)^{1/3}$, which are consistent with Eqs.~(\ref{dmu_genwall}) and (\ref{x0_genwall}), respectively, for $\nu=2$, as expected.

%A different class of models is represented by periodic structures, in which case another length is involved. Here we consider sinusoidally-shaped slits described by the function $\psi(x)=-A\cos(kx)$

%(with $A<L/2$), which beside capillary condensation allow possibly for a formation of an array of liquid films bridging the adjacent parts of the walls. Let us first contemplate the behaviour of the

%bridging transition upon flattening the walls as $k\to0$, which in this case leads to a divergence of the bridging film widths as

% \bb

%x_0\sim k^{-\frac{2}{3}}+{\cal{O}}(1)\,, \label{x0_sin}

%  \ee

%i.e. more slowly than the period itself, meaning that in the re-scaled, dimensionless units, $x\to kx$, the film thickness shrinks to zero.

% Here, however, the bridging transition differs qualitatively from the previous case in

%two ways. Firstly, it precedes capillary condensation, such that

% \bb

% \delta\mu_B=\delta\mu_{\rm cc}(L)\left[1+\frac{2A}{L}+{\cal{O}}(k^{\frac{2}{3}})\right]\,,\label{dmu_sin}

% \ee

%located effectively as capillary condensation in a slit of width $L-2A$, provided the period of the walls is large (compared to $A$ or $L$).

%Secondly, in sinusoidally-shaped slits bridging transition only occurs when their width is below a certain boundary $L_M$, scaling as

% \bb

% L_M\propto\sqrt{\frac{A}{k}}\,,\;\;\;A\to0\,.  \label{LM}

% \ee

%Upon decreasing $k$,

These macroscopic predictions have been tested by a microscopic classical density functional theory (DFT)  \cite{evans79}. Within DFT, one minimizes the grand potential functional
\begin{equation}
\Omega[\rho]={\cal{F}}[\rho]-\int(\mu-V(\rr))\rho(\rr)\dr\,,
\end{equation}
to determine the equilibrium density profile $\rhor$ of the fluid particles and the thermodynamic free energy of the system.  Here, $V(\rr)$ is the external potential arising from the confining walls
and ${\cal{F}}[\rho]$ is the intrinsic free energy functional modelling the contribution from the fluid-fluid interactions. The latter is approximated by combining Rosenfeld's fundamental measure
theory \cite{ros} describing accurately packing effects due to repulsive interactions, with a mean-field treatment of the attractive part of the inter-atomic interaction modelled by a truncated
Lennard-Jones (LJ) potential whose parameters, $\sigma$ and $\varepsilon$, are used as respective length and energy units \cite{our_sin}.

 \begin{figure}
 \includegraphics[width=\linewidth]{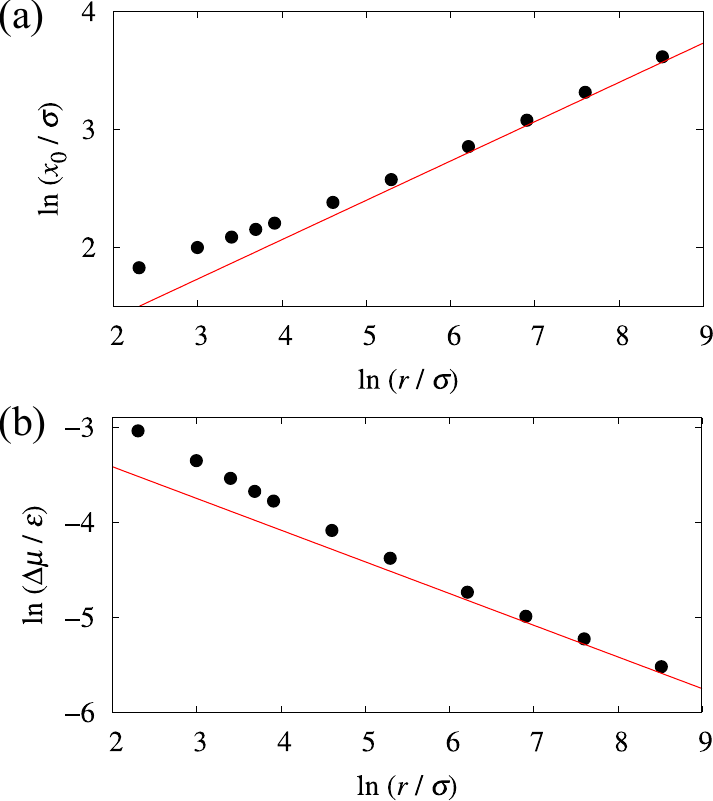}
\caption{Log-log plots showing DFT results for bridging transition between two parallel walls of circular cross-section with a radius $r$ separated by a distance $L=10\,\sigma$.  Panel a) displays
the growth of the bridging film and panel b) the chemical potential offset in the chemical potential between bridging transition and capillary condensation $\mu^{\rm slit}_{\rm cc}(L)$ upon
increasing $r$. Also shown are the expected asymptotic behaviour as given by Eqs~(\ref{x0_cyl}) and (\ref{dmu_cyl}), respectively.} \label{fig_cyl}
\end{figure}

 \begin{figure}
\includegraphics[width=\linewidth]{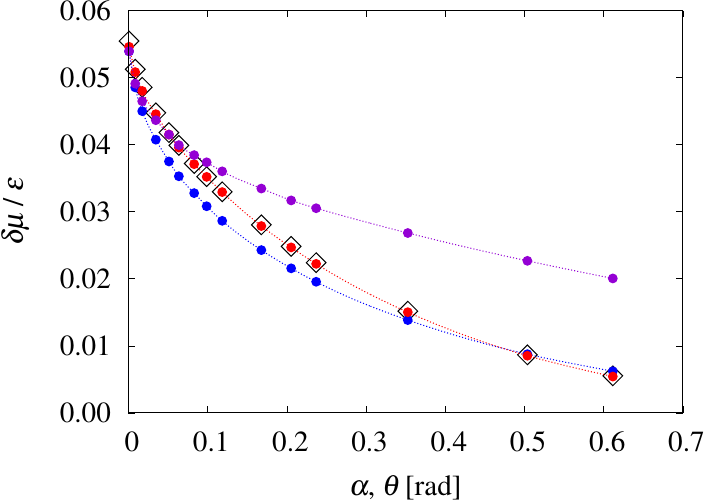}
 \caption{DFT results showing a comparison between the location of bridging transition $\delta\mu_B^\alpha$ (squares) inside a double-wedge slit formed of completely wet walls ($\theta=0$) with the
tilt angle $\alpha$ and the closest distance $L=10\,\sigma$  and capillary condensation $\delta\mu^{\rm slit}_{\rm cc}(\tilde{L})$ in a parallel infinite slit formed of partially wet walls with the
contact angle $\theta=\alpha$. The effective slit width $\tilde{L}$ was determined by Eq.~(\ref{leff}), with $x_0$ given by Eq.~(\ref{x0_wedge}) (blue circles) and  directly from DFT (red circles).
Also shown are the DFT results for  $\delta\mu^{\rm slit}_{\rm cc}(\tilde{L})$  with  $\tilde{L}$ given by the asymptotic relation (\ref{leff_asm}) (purple circles).} \label{fig_cov}
\end{figure}

%\begin{figure*}

 %\includegraphics[width=\linewidth]{fig3}

%\caption{Log-log plots showing a decrease of the chemical potential shift $\Delta\mu$ between capillary condensation and bridging transitions inside the power-law shaped walls (Eq.~(\ref{gen_wall}))

%with a) $\nu=1/2$, b) $\nu=1$, and c) $\nu=2$ upon reducing the parameter $\alpha$. In all the cases the minimum distance between the walls is $L=10\,\sigma$. The symbols represent the DFT results,

%while the straight lines have the slopes corresponding, respectively, to the expected values of the critical exponents $2/3$, $1/2$, and $1/3$. Note that in cases a) and b) the asymptotic predictions

%given by Eq.~(\ref{dmu_genwall}) approach the DFT results from below, while in case c) from above, in line with our expectations.}

%\end{figure*}

In Fig.~\ref{fig2} we compare the predictions (\ref{dmu_genwall}) and (\ref{x0_genwall}) with our DFT numerical results for  square-root ($\nu=1/2$), linear ($\nu=1$), and parabolic ($\nu=2$) wall
geometries. In all the cases, the DFT results for the growth of the bridging film thickness (as given by $x_0$) converge  upon reducing the parameter $\alpha$ to the expected asymptotic behaviour
with the appropriate value of the critical exponent $\beta_\alpha$. It is well seen that the growth of the film becomes slower as the exponent $\nu$ is increased and that the asymptotic lines are
always approached from above, as expected. Also verified is the asymptotic rate of decline in $\Delta\mu$ characterizing the shift between the location of bridging transition and capillary
condensation in the corresponding slit. Here, however, in line with the predictions, the manner in which the given asymptote is approached depends on the specific geometry, such that the convergence
is from below for $\nu=1/2$ but from above for $\nu=2$, in which case the convergence is the slowest. Quantitatively similar to the latter case is the asymptotic behaviour of bridging transitions
between a pair of walls of a circular cross-section of radius $r$. As demonstrated in Fig.~\ref{fig_cyl}, the DFT results exhibit the power-laws  predicted by Eqs.~(\ref{dmu_cyl}) and (\ref{x0_cyl})
for sufficiently large values of $r$, with both asymptotes being approached from above, as expected.

Finally, in Fig.~\ref{fig_cov} we test the covariance between bridging transition inside a double-wedge and capillary condensation in an infinite slit, as given by Eq.~(\ref{cov}). To this end, we
compared the DFT results for the location of bridging transition in the double-wedge formed of completely wet walls ($\theta=0$) tilted by angle $\alpha$ relative to the horizontal, with DFT results
for capillary condensation inside a slit formed of \emph{partially} wet walls with the contact angle $\theta=\alpha$ and width $\tilde{L}$, which was determined: (i) analytically, using
Eqs.~(\ref{x0_wedge}) and (\ref{leff}), (ii) analytically, using the asymptotic relation (\ref{leff_asm}), and (iii) semi-analytically, from Eq.~(\ref{leff}) with $x_0$ determined from DFT density
profiles for the double-wedge. The comparison shows a very reasonable agreement for (i) and a perfect agreement for (iii), while the asymptotic form for $\tilde{L}$ is shown to be accurate for
$\alpha\lesssim 0.1$.

In summary, we have studied bridging transitions between two non-planar surfaces. We have shown that the problem can be solved effectively by an appropriate mapping of the system to a much simpler
one formed of a pair of parallel plates using the newly generalized Kelvin equation. The derivation of the equation extends the concept of the edge contact angle, which can be usefully utilized even
for systems where no edges are present. In the second part of this paper, we studied the asymptotic behaviour of bridging transitions and their relation to capillary condensation for a class of
fundamental walls geometries. We have shown that gradual flattening of the confining walls leads to a sort of critical phenomenon characterized by a diverging growth of the bridging film. Associated
geometry-dependent critical exponents were determined and a covariance law  revealing a relation between the geometric and Young's contact angle for wedge-like structures was found. All the
analytical predictions have been verified by a microscopic density functional theory whose results not only support the anticipated asymptotic behaviour of the bridging transitions but also the
expected way at which the asymptotes are approached.

Natural extensions of this study include the analysis of bridging transitions between walls whose shape is described by a function which is not necessarily differentiable, i.e. include cusps. Here,
the solution of Eqs.~(\ref{bridging}) and (\ref{thetae2}) for $x_0$ (specifying the location of the bridging film boundary) must be compared with the one for a state pertinent to a meniscus pinned to
a nearby edge to find a state of the global free-energy minimum; this requires solving the Kelvin-like equation but now for a fixed value of $x_0$ and unknown $\theta_e$. Further but straightforward
modifications are required for a description of bridging transition between unlike walls. Of great interest would also be an analysis of an interplay between bridging and other confinement-induced
phenomena leading to local condensation, such as wedge or groove filling. Finally, it would be desirable to extend the current study by considering surfaces possessing axial, rather than translation
symmetry and also to account for the effect of wetting layers adsorbed at the walls. We will attempt to address these tasks within our future work.

We are grateful to prof. A. O. Parry for helpful discussions.
This work was financially supported by the Czech Science Foundation, Project No. 21-27338S.


\begin{thebibliography}{99}


\bibitem{RW}

 J. S. Rowlinson and B. Widom, {\it Molecular Theory of Capillarity} (Oxford University Press, Oxford, 1982).


 \bibitem{henderson92}

 D. Henderson, {\it Fundamentals of Inhomoheneous Fluids} (Marcel Dekker, New York, (1992).


 \bibitem{binder2008}

K. Binder, J. Horbach, R. L. C. Vink, and A. De Virgiliis, Soft Matter {\bf 4}, 1555 (2008).




\bibitem{NF81}

M. E. Fisher and H. Nakanishi, J. Chem. Phys. {\bf 75}, 5857 (1981)


\bibitem{NF83}

H. Nakanishi and M. E. Fisher, J. Chem. Phys. {\bf 78}, 3279 (1983).


\bibitem{parry90}

A. O. Parry and R. Evans, Phys. Rev. Lett. {\bf 64}, 439 (1990).


\bibitem{gelb}

L. D. Gelb, K. E. Gubbins, R. Radhakrishnan, and M. Sliwinska-Bartkowiak, Rep. Prog. Phys. {\bf 62}, 1573 (1999).




\bibitem{gregg}

 S. J. Gregg and K. S. W. Sing, {\it Adsorption, Surface Area and Porosity}, 2nd edition, Academic Press, New York, (1982).





\bibitem{evans84}

 R. Evans and P. Tarazona, Phys. Rev. Lett. {\bf 52}, 557 (1984).


 \bibitem{evans85}

 R. Evans and U. Marini Bettolo Marconi, Phys. Rev. A 32, 3817 (1985).


\bibitem{evans90}

 R. Evans, J. Phys.: Condens. Matter {\bf 2}, 8989 (1990).




\bibitem{rocken96}

P. Rock\"{o}n and P. Tarazona, J. Chem. Phys. {\bf 105}, 2034 (1996).


\bibitem{rejmer99}

K. Rejmer, S. Dietrich, and M. Napi\'orkowski, Phys. Rev. E {\bf 60}, 4027 (1999).



\bibitem{parry99}

 A. O. Parry, C. Rasc\'on, and A. J. Wood, Phys. Rev. Lett. {\bf 83}, 5535 (1999).


\bibitem{quere02}

D. Qu\' er\'e, Physica A {\bf 313}, 32 (2002).


\bibitem{bruschi02}

L. Bruschi, A. Carlin, and G. Mistura, Phys. Rev. Lett. {\bf 89}, 166101 (2002).


\bibitem{gang}

O. Gang, K. J. Alvine, M. Fukuto, P. S. Pershan, C. T. Black, and B. M. Ocko, Phys. Rev. Lett. {\bf 95}, 217801 (2005).


\bibitem{tas07}

M. Tasinkevych and S. Dietrich, Eur. Phys. J. E {\bf 23}, 117 (2007).


\bibitem{quere08}

D. Qu\' er\'e, Annu. Rev. Mater. Res. {\bf 38}, 71 (2008).


\bibitem{bruschi09}

L. Bruschi and G. Mistura, J. Low Temp. Phys. {\bf 157}, 206 (2009).


\bibitem{hofmann}

 T. Hofmann, M. Tasinkevych, A. Checco, E. Dobisz, S. Dietrich, and B. M. Ocko, Phys. Rev. Lett. {\bf 104}, 106102 (2010).


\bibitem{mistura13}

G. Mistura, A. Pozzato, G. Grenci, L. Bruschi, and M. Tormen, Nat. Commun. {\bf 4}, 2966 (2013).


\bibitem{pos22}

 M. Posp\'\i\v sil, A. O. Parry, and A. Malijevsk\'y, Phys. Rev. E {\bf 105}, 064801 (2022).




\bibitem{mal21}

A. Malijevsk\'y and A. O. Parry, Phys. Rev. Lett. {\bf 127}, 115703 (2021); Phys. Rev. E {\bf 104}, 044801 (2021).



\bibitem{our_slit}

A. Malijevsk\'y, A. O. Parry, and M. Posp\'\i\v sil, Phys. Rev. E {\bf 96}, 020801(R) (2017).




\bibitem{wenzel}

R. N. Wenzel, Ind. Eng. Chem. {\bf 28}, 988 (1936).

\bibitem{unbending}

 C. Rasc\'on, A. O. Parry, and A. Sartori, Phys. Rev. E {\bf 59}, 5697 (1999).

\bibitem{manuel}
 \'A. Rodr\'\i guez-Rivas, J. Galv\'an and J. M. Romero-Enrique, J. Phys: Condens. Matter {\bf 27}, 035101 (2015).


\bibitem{rascon2000}
C. Rasc\'on and A. O. Parry, Nature {\bf 407}, 986 (2000).



\bibitem{posp_22}
M. Posp\'\i\v sil and A. Malijevsk\'y, Phys. Rev. E {\bf 106}, 024801 (2022).


\bibitem{evans79}

R. Evans, Adv. Phys. {\bf 28}, 143 (1979).


\bibitem{ros}

Y. Rosenfeld,  Phys. Rev. Lett. {\bf 63}, 980 (1989).


\bibitem{our_sin}

For details regarding the DFT model see  \cite{posp_22}.



\end{thebibliography}
\end{document}